\documentclass[
    ,final            
    ,numberedheadings 
    ,cmfonts          
  ]
  {aipproc}

\layoutstyle{6x9}

\newcommand \be{\begin{equation}} %
\newcommand \ee{\end{equation}} %
\newcommand \bea{\begin{eqnarray}} %
\newcommand \eea{\end{eqnarray}} %
\newcommand{\degree}{\ensuremath{^\circ}}
\usepackage{amssymb} 

\begin{document}

\title{Axions and polarisation of quasars}

\classification{
		14.80.Mz,
		95.30.Gv,
		98.54.Aj,
		98.65.Dx
	       }

\keywords      {
		axion,
		polarisation,
		quasars,
		large-scale structure of the Universe
	       }

\author{A.~Payez\footnote{IISN Research Fellow}\ }{
  address={D\'epartement d'Astrophysique, de G\'eophysique et d'Oc\'eanographie, Universit\'e de  Li\`ege, B5a, Sart-Tilman, B-4000 Li\`ege, Belgium}
}

\author{J.R.~Cudell}{
  address={D\'epartement d'Astrophysique, de G\'eophysique et d'Oc\'eanographie, Universit\'e de  Li\`ege, B5a, Sart-Tilman, B-4000 Li\`ege, Belgium}
}

\author{D.~Hutsem\'ekers\footnote{Senior Research Associate FNRS}\ }{
  address={D\'epartement d'Astrophysique, de G\'eophysique et d'Oc\'eanographie, Universit\'e de  Li\`ege, B5a, Sart-Tilman, B-4000 Li\`ege, Belgium},
altaddress={E-mails: A.Payez@ulg.ac.be, JR.Cudell@ulg.ac.be, Hutsemekers@astro.ulg.ac.be}
}
\begin{abstract}

We present results showing that, thanks to axion-photon mixing in external magnetic fields, it is actually possible to produce an effect similar to the one needed to explain the large-scale coherent orientations of quasar polarisation vectors in visible light that have been observed in some regions of the sky.

\end{abstract}

\maketitle


\section{Introduction}

Axions are thirty years old. Their hypothetical existence has indeed been proposed, simultaneously by Weinberg and Wilczek in 1978~\cite{w_w}, as a consequence of the Peccei--Quinn solution to the strong CP problem~\cite{P_Q}, which was at that time ---and still is--- the preferred one.

Thirty years have passed and axions are yet to be observed. Actually, they are thought nowadays to be very weakly interacting particles, with a mass below the eV, making their detection quite challenging.\footnote{This is why they are sometimes called ``invisible axions''.} Of course, some people have tried, and are trying, to detect them ---in particular, the recent, now-retracted~\cite{PVLAS}, PVLAS signal, interpreted as the effect of an unusual axion, stirred a lot of interest--- but, until now, no one managed to see such particles.

If they do exist, axions must interact with light. This last statement is especially interesting as this interaction would help understand the recent observation, in visible light, of alignments over huge distances ($\sim$~1~Gpc) of polarisation vectors of quasars\footnote{In this text, we will use the words \emph{quasars} and \emph{AGN} (Active Galactic Nuclei) interchangeably.}~\cite{hutsemekers}. In the following, we will thus investigate axion-photon mixing in external magnetic fields, a subject which has already been discussed by several authors~\cite{axionphotonmixing} ---and, very recently, by Adler \emph{et al.}~\cite{adler}--- and we will concentrate on this particular context of astrophysics and will show that (very light) axions can produce effects on light polarisation likely to explain the observations.

\section{Coherent orientation of quasar polarisation vectors}

\begin{figure}[htbp]
        	\resizebox{12.5cm}{!}{\rotatebox{-90}{
					\includegraphics[width=\textwidth]{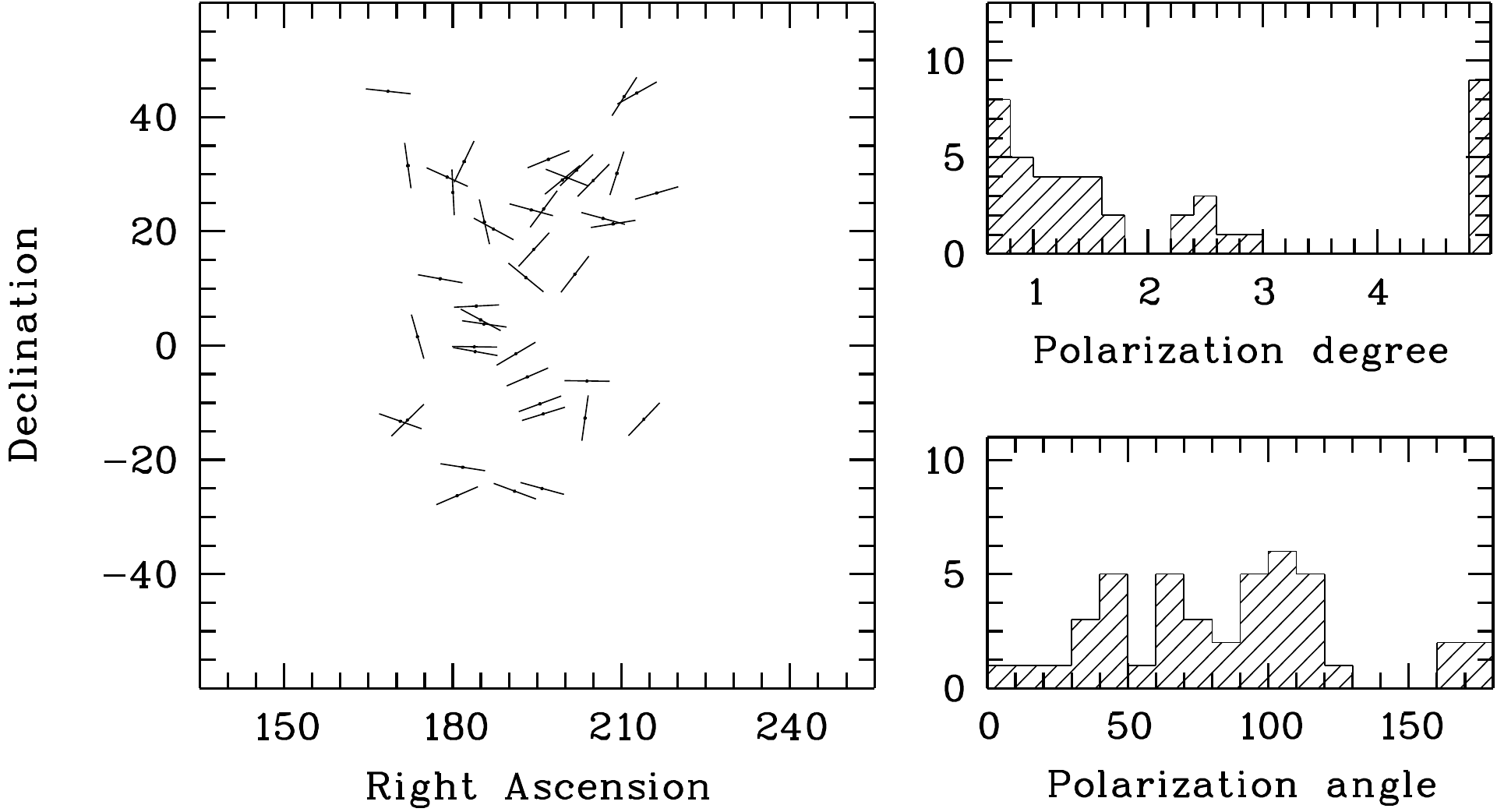}
        	}}
        	\caption{Map of quasar polarisation vectors: the coordinates are in degrees and all these objects have a redshift lying between 0 and 1~\cite{hutsemekers}. On the upper right-hand side diagram, we see the number of objects for a certain value of the degree of polarisation, in percents; the last bin being for objects with $p\geq 5$\%. The diagram below shows the distribution of polarisation angles (defined between 0\degree and 180\degree) for this particular region, their mean value being $\bar{\theta}=79\degree$.}
	\label{lowz}
\end{figure}

\begin{figure}[htbp]
        	\resizebox{12.5cm}{!}{\rotatebox{-90}{
					\includegraphics[width=\textwidth]{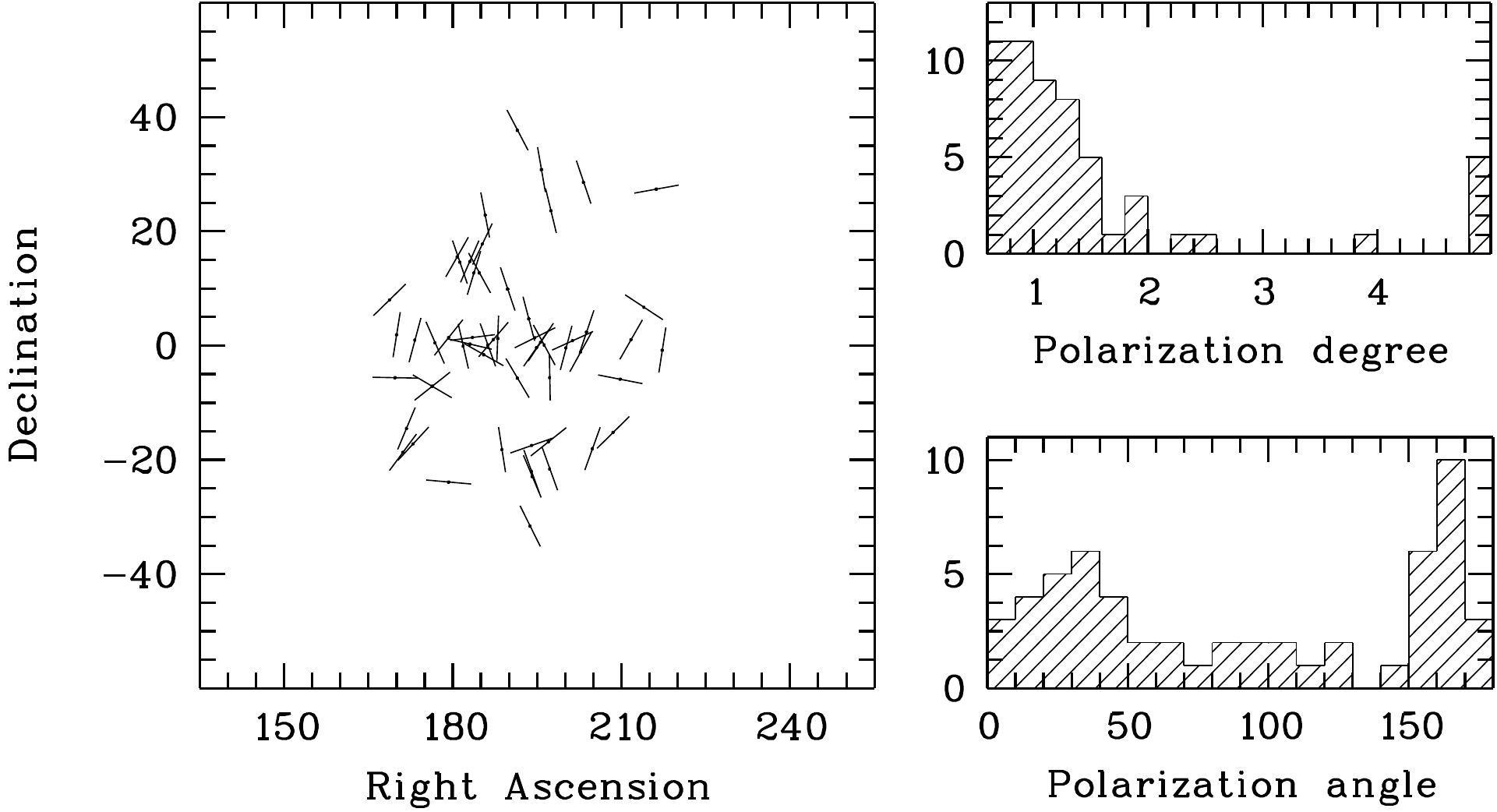}
        	}}
        	\caption{Map of quasar polarisation vectors: same region as above, except that the redshifts are now lying between 1 and 2.3~\cite{hutsemekers}. The mean polarisation angle for this region is $\bar{\theta}=8\degree$.}
	\label{highz}
\end{figure}

We shall start with a quick review of the alignment effect, namely the observation that polarisation vectors for visible light coming from different quasars tend to be coherently oriented over cosmological scales in some regions of the sky. The regions under discussion are located at both low (Fig.~\ref{lowz}) and high (Fig.~\ref{highz}) redshifts and have different preferred orientations of quasar polarisation for similar lines of sight.

This effect has been confirmed on a sample of 355 quasar linear polarisation measurements (half of them measured by~\cite{hutsemekers}, and half coming from the literature), chosen because of their accuracy and because they are away from the Galactic plane: different statistical tests give a probability for coincidental alignments from 0.1 to 0.01\%.

We will not discuss further these observations here (for details, see~\cite{hutsemekers}), however, let us emphasise that, as the mean directions of polarisation are not similar at different redshifts along similar lines of sight, this correlation effect is not likely to be explained in terms of local effects.
This is why axion-like particles come into play, as they would provide a mechanism affecting light during its propagation.

\section{Axion--photon mixing}\label{axion_photon_mixing}

Axions, if they exist, are pseudoscalar particles, like the $\pi^0$, and, as such, they couple with photons and are subject to what is called, in the pion case, the Primakoff effect~\cite{Primakoff}: there is a probability for a photon travelling through an external magnetic (or electric) field to give rise to a pseudoscalar\footnote{The main difference being that the pion is much more massive and interacts more strongly than the axion, the latter being almost stable.} and, similarly, for the reverse process to take place. The polarisation for the light undergoing this process ought to be parallel to the external magnetic field, the perpendicular component being insensitive to axion effects, as we shall see in section~\ref{axions_and_light_polarisation}.

\subsection{Dichroism}

A first consequence of this axion--photon mixing is a dichroism, \textit{i.e.} a selective absorption of one direction of polarisation, which, in turn, leads to the appearance of a slight linear polarisation for non-fully polarised beams\footnote{Which can always be decomposed as the sum of two orthogonal linearly polarised ones.}. We thus have a mechanism likely to create additional polarisation.

Side effects of this dichroism are a possible\footnote{If the initial beam is fully polarised in the external magnetic field direction, we will not have a rotation. Note that a rotation of the mean direction of polarisation in the case of quasars might have been observed, however more data should be taken to confirm this~\cite{hutsemekers}.} rotation of the polarisation plane and, of course, a modification of the intensity of light (a loss, when starting with a photon beam, or a gain, in the case of a ``light-shining through wall'' experiment).

\subsection{Birefringence}

Another consequence is birefringence induced by axion--photon mixing. Birefringence, \textit{i.e.} the fact that one direction of polarisation is travelling more slowly than the other, can be induced by double Primakoff effect (or, similarly, virtual axions): if this happens, the polarisation parallel to the external magnetic field picks up, for some time, the mass of the axion\footnote{This mass could, actually, be higher or smaller, in the case of extremely light axion-like particles (ALPs), than the effective mass ---the plasma frequency--- acquired by the photon in the intergalactic medium (which is not strictly the vacuum), $\omega_p\sim\mathcal{O}(10^{-14}~\textrm{eV})$. In the following, we will consider ALPs with masses of that order.}, inducing a phase shift between the parallel and perpendicular components of the beam, leading finally to the generation of elliptical polarisation in the most general case.

\section{Axions and light polarisation}\label{axions_and_light_polarisation}

Using natural units and the Heaviside--Lorentz convention and following~\cite{axionphotonmixing}, we can write the axion--photon lagrangian density as

\be \mathcal{L} =  \frac{1}{2}\ (\partial_{\mu}\phi) (\partial^{\mu}\phi) - \frac{1}{2}\ m_{\phi}^2\phi^2 - \frac{1}{4}\ F_{\mu\nu} F^{\mu\nu}
         - j_{\mu} A^{\mu} + \frac{1}{4}\ g \phi F_{\mu\nu}\widetilde{F}^{\mu\nu}, \label{eq:lagrangien}
\ee
where we recognise the Klein--Gordon and Maxwell parts, followed by a new term, $\frac{1}{4}\ g \phi F_{\mu\nu}\widetilde{F}^{\mu\nu}$, responsible for the mixing of axions with photons.
One can then obtain the equations of motion for the fields, adding a plasma frequency $\omega_p$ in the discussion for completeness:
\begin{eqnarray}\left\{\begin{array}{llll}
\Box E_{\perp} +~\! \omega_p^2 E_{\perp} = ~0\\
\Box E_{\parallel}~\!+~\! \omega_p^2 E_{\parallel} ~\!= ~g B^{ext}_T \partial^2_t\phi\\
\Box \phi~~~\!\!+~\! m^2\phi ~~\!= - gE_{\parallel}B^{ext}_T
\label{eq_mvt}\end{array} \right.,\end{eqnarray}
where we have assumed that the terms involving the longitudinal part of the electric field are negligible, that the process will take place at large distances compared to the size of a quasar (so we put $\rho$ and $\vec{j}$ = 0), and, finally, that the magnetic field of a single photon is much smaller than the external magnetic field $B^{ext}$. All vectors here are defined in the plane perpendicular to the beam; $E_{\perp}$ (resp. $E_{\parallel}$) is the component perpendicular (resp. parallel) to $B^{ext}_T$, the transverse part of the external magnetic field.

Simply looking at \eqref{eq_mvt}, it is immediate to see that, indeed, as we discussed when we talked about the Primakoff effect, only the part of the electric field parallel to $B^{ext}$ is coupled to the axion field, whereas the perpendicular part remains unaffected.
\\

\begin{figure}[htbp]
	\includegraphics[width=\textwidth]{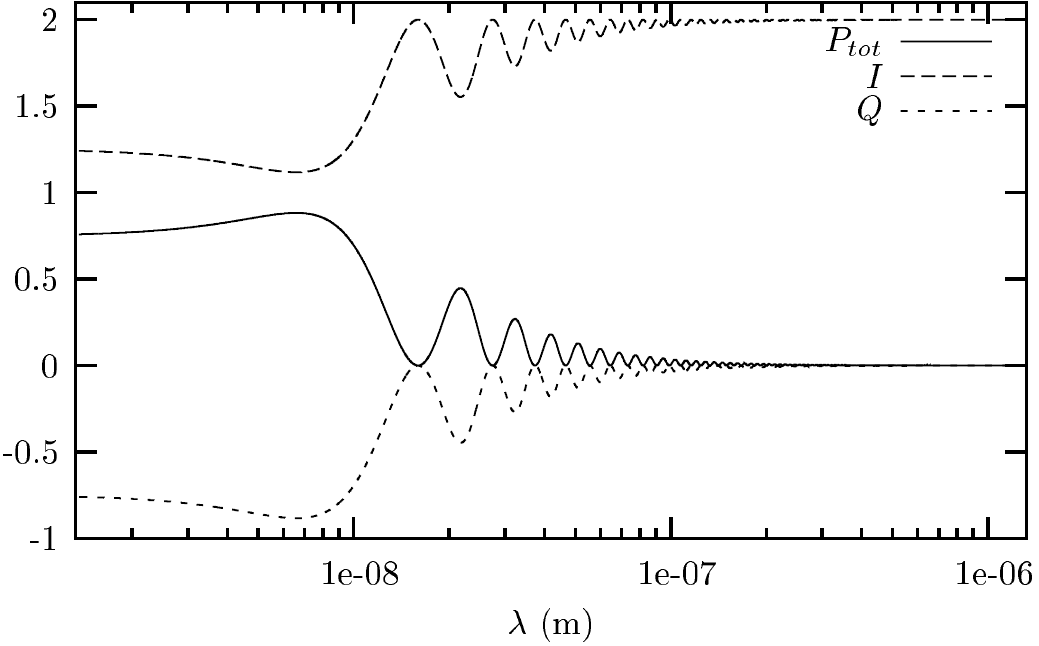}
        	\caption{The polarisation, $P_{tot}$, $I$ and $Q$, for different wavelengths, at the end of a 10~Mpc magnetic field zone in the case of initially unpolarised light beams, using, as parameters: $\omega_p = 3.6452~10^{-14}$~eV,  $m_{\phi} = 1.15271~10^{-14}$~eV, $gB^{ext}_T = 1.46615$~Mpc$^{-1}$ (e.g. $g= 4.8~10^{-12}$~GeV$^{-1}$ and $B^{ext}_T = 0.1~\mu$G). The parameters $U$ and $V$, not shown here, are zero at all wavelengths.}
	\label{ini_npol}
\end{figure}

\begin{figure}[htbp]
	\includegraphics[width=\textwidth]{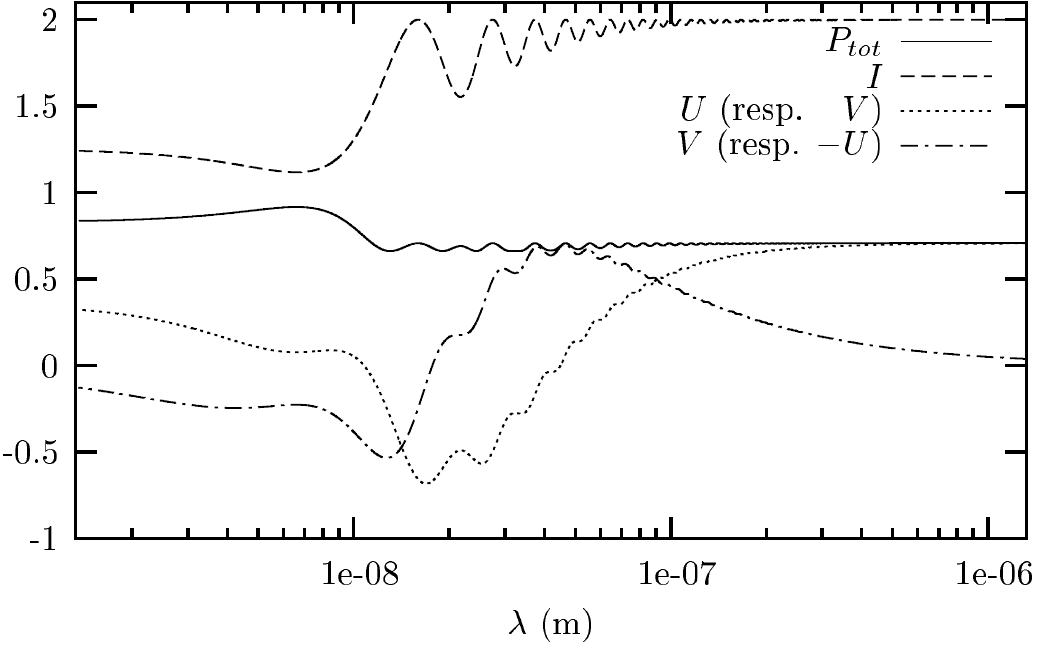}
        	\caption{The polarisation, $P_{tot}$, $I$, $U$ and $V$, for different wavelengths, at the end of a 10~Mpc magnetic field zone in the case of initially polarised light beams, using the same parameters as in Fig.~\ref{ini_npol}. The linear polarisation, $Q$, not shown here for readability, and the intensity, $I$, actually have still exactly the same behaviour as in the unpolarised case.}
	\label{ini_pol}
\end{figure}

However, to make dichroism and birefringence explicit, one should compute quantities used to describe the polarisation state of the light beam: the Stokes parameters. Several notations for these parameters exist in the literature; here, we will use $I$ for the intensity, $Q$ and $U$ to describe the linear polarisation and $V$, the circular polarisation:\footnote{To gain more insight into what these parameters really mean, the reader is referred to \cite{Jackson}.}
                \bea
                \left\{
                    \begin{array}{llll}
                        I(x)  &=& E_{\parallel}(x) E^*_{\parallel}(x) + E_{\perp}(x) E^*_{\perp}(x)\\
                        Q(x) &=& E_{\parallel}(x) E^*_{\parallel}(x) - E_{\perp}(x) E^*_{\perp}(x)\\
                        U(x) &=& E^*_{\parallel}(x) E_{\perp}(x) + E_{\parallel}(x) E^*_{\perp}(x)\\
                        V(x) &=& i(-E^*_{\parallel}(x) E_{\perp}(x)  + E_{\parallel}(x) E^*_{\perp}(x))
                    \end{array} \right..
                    \label{eq:Stokes}
                \eea\\
Using the solutions of equations~\eqref{eq_mvt} in~\eqref{eq:Stokes}, we have obtained the values, for different values of the wavelength, of the Stokes parameters and of the polarisation, $P_{tot}=\sqrt{Q^2 + U^2 + V^2}$, at the end of a 10~Mpc zone of constant magnetic field, using plane waves: firstly, for an initially unpolarised beam (Fig.~\ref{ini_npol}) and, then, for an initially polarised one (Fig.~\ref{ini_pol}), both in the absence of an initial axion field, $\phi(0) = 0$. Of course, unpolarised light cannot be understood without the concept of wave packets; here, by using plane wave solutions, we actually chose a very special kind of wave packet, \textit{i.e.} a sum of plane waves with the same frequency but with different direction of polarisation, giving either an overall non-polarised beam or an initially polarised one (the Stokes parameters are here to be understood as a mean over the data acquisition time).
\\

If we look at Fig.~\ref{ini_npol}, we see that, while we have considered light beams, of different wavelengths, all being initially unpolarised ---that is, $I(0) = Q(0) = U(0) = V(0) = 0$---,  we end up with a spontaneous appearance of polarisation at some wavelengths after they have travelled 10~Mpc through the external magnetic field (let us insist that these figures do not show the evolution of the Stokes parameters with distance, whose behaviour is oscillatory).

The polarisation mimics the evolution of the parameter Q (we indeed have $P_{tot} = \sqrt{Q^2 + U^2+ V^2}=|Q|$) and mirrors that of $I$. 
This is actually the signature of dichroism. What happens is that we have a depletion of photons with polarisation parallel to $B^{ext}_T$, leading to a loss of intensity and to a modification of the $Q$ parameter (see equation~\eqref{eq:Stokes}) by the same amount. This, in turn, will lead to a net appearance of polarisation.

What can also be noticed on this figure is an oscillatory behaviour, with increasing amplitude from large to small values of wavelength, ending with a plateau (with no more oscillations), the argument of the oscillatory function being then constant ---for other values of the parameters, this plateau will take other values.
\\

If we now examine the second figure, we find other interesting features of this axion--photon mixing in magnetic fields, concerning the linear polarisation $U$ and the circular polarisation $V$.

First of all, we have noticed that there is some kind of symmetry in the law of evolution of the parameters $U$ and $V$, in the special case where one of the two is initially equal to zero. An illustration of this has been represented on Fig.~\ref{ini_pol}, which can actually be interpreted as the case of a light beam either with $U(0)=\frac{1}{\sqrt{2}}$ and $V(0)=0$ or with $U(0)=0$ and $V(0)=\frac{1}{\sqrt{2}}$; while we give the other Stokes parameters the same values as for Fig.~\ref{ini_npol}. Writing explicitly the expressions for $U(x)$ and $V(x)$ using the solutions of Eq. \eqref{eq_mvt}, one can see that the law of evolution of $U$ (resp. of $V$) in the first case is exactly the one for $V$ (resp. for $-U$) in the second one.
This is in fact a signature of birefringence, which can be seen as due to some retarder. It is indeed well-known that using retarders, one can produce a circularly polarised beam, starting with a linearly polarised one and \textit{vice-versa}.

This being, maybe the most striking observation that can be made looking at Fig.~\ref{ini_npol} is that, starting with a very general linearly polarised light, we will always end up with circular polarisation ---at least in our particular case of plane waves.
This is related to birefringence, once again: when birefringence occurs, some of the photons polarised parallel to $B^{ext}$ are slowed down (or speeded up, depending on the axion mass), phase shifted, leading to the appearance of some circular polarisation, by definition. The reverse statement is equally true as the requirements for pure linearly (the components of the electric field oscillate strictly in phase) and for circularly (phase shifted exactly by $\frac{\pi}{2} + k\pi$) polarised states are both very demanding and will be disturbed by birefringence.

Finally, there is another interesting feature related to these two parameters, not explicit on Fig.~\ref{ini_pol}, in their contribution to the polarisation: one can obtain that this quantity, $U^2+V^2$, can simply be written as $U^2(0)+ V^2(0)$ times some function of the distance and of the other parameters. This shows that, in the case of $U$ and $V$, the only quantity that matters for the polarisation is, in fact, the sum of their initial values squared, notwithstanding the details of their individual initial values --- this confirms a close connection between $U$ and $V$ in these processes.

\begin{center}
	\begin{figure}[htbp]
        	\resizebox{12.5cm}{!}{\rotatebox{90}{
	\includegraphics{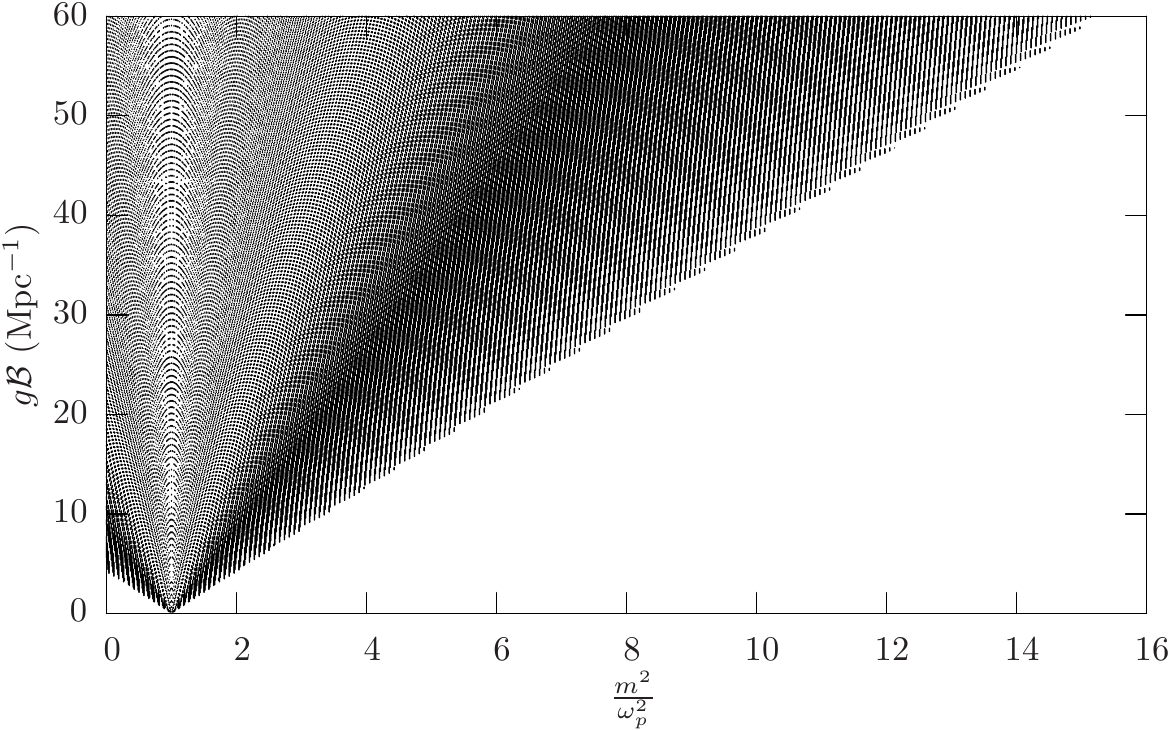}
	\label{eventail}
        	}}
        	\caption{Parameter space, for a 500~nm initially unpolarised light beam, able to give an additional polarisation between 0.005 and 0.02, at the end of a 10~Mpc magnetic field zone ---this has been obtained using a fixed value for the plasma frequency which is typical in clusters and superclusters ($\omega_p = 2.6452~10^{-14}$~eV).}
	\end{figure}
\end{center}

Now that we have seen that the coupling of axions to photons can induce a modification of the polarisation state of a light beam, we can ask whether this mixing is able to explain the observations of quasar polarisation vectors, which require a significant additional linear polarisation estimated as
\[0.005\leq\frac{\sqrt{Q^2 + U^2}}{I}\leq 0.02.\]
Considering again a very light axion-like particle and using our very simple case of summed up plane waves travelling through a constant external magnetic field of length 10~Mpc, we have drawn the space of parameters (see Fig.~\ref{eventail}) which fulfil this requirement for the observed wavelength. The depletion of points in the vicinity of $m\approx\omega_p$ is explained by an enhancement of the effect (the amplitude of the linear degree of polarisation goes to one, the beam being in this case fully linearly polarised) while their absence in regions where the two masses are very different or where the product $gB^{ext}$ is small is explained by the weakness of the effect.

\section{Conclusion}

We have discussed axion-photon mixing in external magnetic fields and their consequences on the polarisation state of light beams undergoing this process and have particularly emphasised the interpretations of our results on the evolution of Stokes parameters in terms of dichroism and birefringence. 

Finally, we have shown that this mechanism is able to provide a modification of the polarisation state likely to explain the large-scale coherent orientations of polarisation vectors of visible light (500~nm) coming from quasars.

However, to obtain these results, we have used typical values for all the parameters, except that we had to consider very light pseudoscalar particles, with a mass similar to the value of the plasma frequency in clusters of galaxies ---the value of $g$ used being consistent with bounds given by experiments on the coupling for such light axion-like particles.


\begin{theacknowledgments}
A.~P. would like to thank the IISN for funding, to acknowledge constructive discussions with C\'edric Lorc\'e on mathematical and physical matters and to thank Sudeep Das and Pankaj Jain for clarifying some issues.
\end{theacknowledgments}


\begin{thebibliography}{9}

\bibitem{w_w}
S. Weinberg, \emph{Phys. Rev. Lett.} \textbf{40} 223--226 (1978); F.~Wilczek, \emph{Phys. Rev. Lett.} \textbf{40} 279--282 (1978).

\bibitem{P_Q}
R.~D.~Peccei and H.~R.~Quinn, \emph{Phys. Rev. Lett.} \textbf{38} 1440--1443 (1977).

\bibitem{PVLAS}
E.~Zavattini \textit{et al.}, [arXiv:hep-ex/0706.3419v1].

\bibitem{hutsemekers}
D.~Hutsem\'ekers, \emph{A\&A} \textbf{332}  410--428 (1998);  D.~Hutsem\'ekers and H.~Lamy, \emph{A\&A} \textbf{367} 381--387 (2001); D.~Sluse \textit{et al.}, \emph{A\&A} \textbf{433} (2005) 757--764; D.~Hutsem\'ekers, R.~Cabanac, H.~Lamy and D.~Sluse, \emph{A\&A} \textbf{441} (2005) 915--930.

\bibitem{axionphotonmixing}
See, for example, G.~Raffelt and L.~Stodolsky, \emph{Phys. Rev. D} \textbf{37} 1237--1249 (1988); S.~Das, P.~Jain, J.~P.~Ralston and R.~Saha, \emph{JCAP} \textbf{0506} 002 1--30 (2005).

\bibitem{adler}
S.~L.~Adler, J.~Gamboa, F.~M\'endez and J.~L\'opez-Sarri\'on, [arXiv:hep-ph/0801.4739v4].

\bibitem{Primakoff}
H.~Primakoff, \emph{Phys. Rev.} \textbf{81} 899 (1951).

\bibitem{Jackson}
J.D.~Jackson, \emph{Classical Electrodynamics}, 2$^{\textrm{\tiny{nd}}}$ edition, Wiley, 1975, p. 273--278.

\end{thebibliography}
\end{document}